# The ensemble effects on adsorption and dehydrogenation of ethylene on PdAu(001) bimetallic surfaces.

Dingwang Yuan[1,2] Xingao Gong[3] and Ruqian Wu[1]

*1. Department of Physics and Astronomy, University of California, Irvine, CA 92697-4575*

*2. ICTS, Chinese Academy of Sciences, Beijing, 100080, China*

*3. Surface Science Laboratory and Department of Physics, Fudan University, Shanghai-200433, China*

Abstract

First-principles calculations are performed to study the adsorption and dehydrogenation of ethylene on PdAu(001) bimetallic surfaces. The activation energies for C−H bond breaking of ethylene strongly depend on the ensemble effect. Particularly, ethylene is unlikely to decompose on the most popular ensembles of PdAu(001), Pd monomers or Pd second neighborhoods, since the Pd-Au bridges are less attractive towards vinyl and the eliminated H atom than the Pd-Pd bridges in the transition state. Correlations are found among the activation energies, reaction energies and adsorption energies, but no obvious relationship exist between the surface activity and the Pd-d band center.

PACS numbers: 68.43.Fg, 71.15. Mb, 71.23.-k, 82.45.Jn

Bimetallic surfaces with tunable chemical properties show an ample potential for electro-catalysis and heterogeneous catalysis applications and, therefore, have attracted broad attention in recent years.[1] For instance, PdAu bimetallic catalysts are highly effective in promoting a variety of reactions such as CO oxidation,[2, 3] $H_2$ oxidation to $H_2O_2$,[4] alcohol oxidation,[5] and vinyl acetate (VA) synthesis.[6] The local chemical properties of constituents are strongly altered from their parent metals by the "ligand effects", which generally describe influences of charge transfer, orbital rehybridization and lattice strain[7, 8]. Extensive experimental and theoretical studies have established clear correlations between the chemical activity of bimetallic surfaces and their electronic features such as core level shift and position of d-band center with respect to the Fermi level.[7,9,10] On the other hand, the "ensemble effects" associated with particular arrangements of the active constituents have received much less attention, even though they are equally or, in some cases, more important towards the reactivity and selectivity of bimetallic catalysts. Goodman's group found that the presence of "Pd monomer pairs" is the primary reason for the remarkable enhancements in vinyl acetate formation rates on PdAu(001). Therefore, it is crucial to attain comprehensive understandings of the ensemble effects for the rational design of robust catalysts. From measurements with scanning tunneling microscope (STM), temperature programmed desorption (TPD) and infrared absorption spectroscopy (IRAS), Behm's[11] and Goodman's group[12] found that the population of the Pd first neighborhoods is extremely small on PdAu(001) and PdAu(111) bimetallic surfaces annealed up to 800 K. This observation was explained in our recent density functional studies and, furthermore, we pointed out the energy preference of forming various Pd second neighborhoods on these surfaces[13].

In this paper, we address the consequence of lack of first Pd neighborhoods on adsorption and dehydrogenation of ethylene on PdAu(001). Many chemical reactions involve the dehydrogenation of ethylene but the mechanism of this reaction is still elusive. Complete dehydrogenation of ethylene occurs easily on pure Pd surfaces, which results in the formations of Pd carbides during VA synthesis on Pd surfaces or large Pd clusters.[14] Although it was reported that the alloying of Au with Pd effectively prevents $PdC_x$ formation,[15] it is unclear if ethylene molecules partially decompose on PdAu. Clear understanding of ethylene dehydrogenation is imperative to appreciating complex reactions such as VA synthesis. It was proposed that VA forms through coupling of acetate and vinyl species on adjacent Pd sites, for which the ethylene dehydrogenation is a critical precursor reaction.[16] The second possibility is that ethylene reacts directly with acetate nucleophile to form an ethyl acetate-like intermediate, followed by β-H elimination. Our calculations found the activation energy very high on the second Pd neighborhoods; and hence refute the first mechanism for VA synthesis.

The calculations are performed using the plane-wave based Vienna ab initio Simulation Package (VASP) [17], in the level of Generalized Gradient Approximation (GGA)[18]. The effects of nuclei and core electrons are represented by ultra-soft pseudopotential.[19] To ensure the numerical convergence, the plane wave expansion has a large energy cutoff of 350 eV, while the reciprocal space is sampled by 5x5x1 k-points in the Monkhorst-Pack[20] grids. The Au(001) surfaces are modeled with a four-layer slab and a 15 Å vacuum. In the lateral plane, we use a c(4x4) supercell, as shown in Fig. 1, so as to reduce the interaction between adjacent adsorbates. Representative Pd distribution patterns in the outmost layer include isolated Pd monomers as well as Pd second and first neighborhoods. We begin with adsorption studies of ethylene, vinyl and H on different ensembles. Positions of all atoms except those in the two bottommost Au layers are fully relaxed according to the calculated atomic forces. The most preferred adsorption sites of ethylene on different ensembles are sketched in Fig.1, for geometries along both the [100] and [110] axes. The minimum energy paths (MEPs) and activation energies for C-H bond breaking are obtained with the Climbing Image Nudged Elastic Band (CI-NEB) approach[21, 22, 23, 24], based on the optimized adsorption geometries for reactants and products.

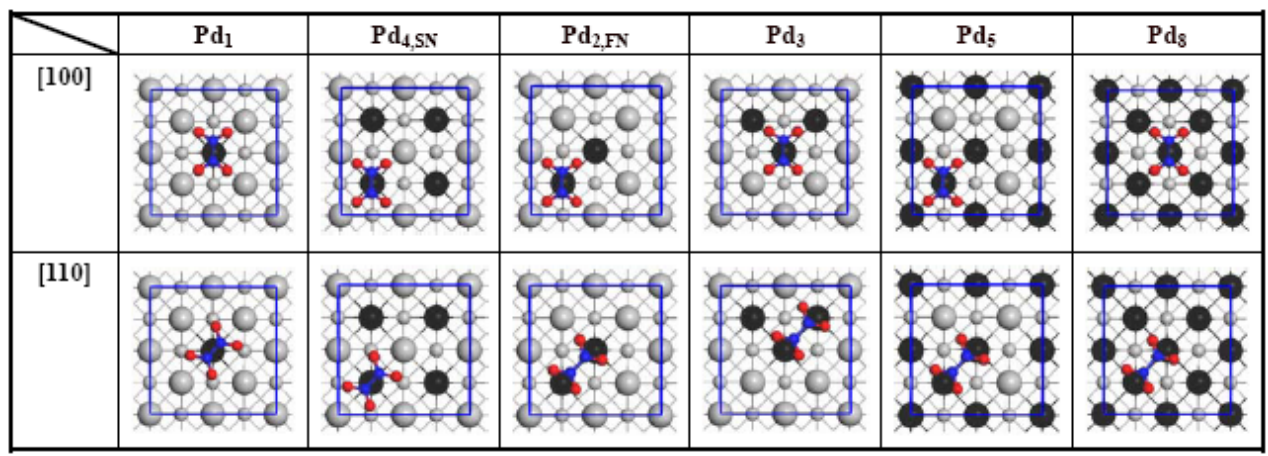

FIG. 1: The unit cells and structures of different Pd ensembles on PdAu(001), along with stable adsorption geometries of an ethylene molecule. The large and small gray balls represent surface and subsurface Au atoms. The black balls are Pd atoms, and small blue and red balls are for C and H atoms, respectively.

The chemical activity of the PdAu surface towards an adsorbate x *(x=ethylene, vinyl and H)* is normally characterized by its adsorption energy

$$E_{ad,x} = -(E_{x/PdAu} - E_{PdAu} - E_x) \qquad (1)$$

where the quantities in parentheses are total energies calculated for different systems. As known,

ethylene takes two stable adsorption geometries on the pure Pd(001) surface, illustrated by different orientations of its C−C bond along either the [100] (π-bonded type) or the [110] (di-σ-bonded) axis. As listed in Table 1, our calculations indicate that the di-σ-bonded geometry is more preferred on all ensembles except $Pd_5$. On the clean Pd(001) surface, the calculated $E_{ad,Ethylene}^{[100]}$ and $E_{ad,Ethylene}^{[110]}$ are 0.82 eV and 0.94 eV, respectively. These data are very close to results of other density functional calculations.[25] It is obvious that ethylene adsorbs stably on all the ensembles of PdAu(001) at room temperature, as was observed experimentally. On an isolated monomer ($Pd_1$) and second neighborhoods ($Pd_{2,SN}$ and $Pd_{4,SN}$), the ethylene molecule takes the top site over Pd and the change in orientations of the C-C bond hardly affect adsorption energies. Unlike CO, which mostly probes properties of a single Pd site, $C_2H_4$ becomes aware of the presence of the second neighbor Pd as manifested by the noticeable increase in $E_{ad,ethylene}$ on the $Pd_1$, $Pd_{2,SN}$ and $Pd_4$ ensembles. Similar to what occurs on the Pd(001) surface, ethylene prefers the di-σ-bonded geometry over the Pd-Pd bridge site. Overall, $E_{ad,ethylene}$ increases monotonically as the number of Pd atoms in the ensemble grows. Particularly, $E_{ad,ethylene}$ on the $Pd_8$ (a complete Pd monolayer) are larger than those on the clean Pd(001) surface, owing to both strain and chemical effects.

TABLE I: The adsorption energies for ethylene ($E_{ad,Ethylene}^{[100]}$ and $E_{ad,Ethylene}^{[110]}$), vinyl ($E_{ad,vinyl}$) and hydrogen ($E_{ad,H}$), the reaction energies ($\Delta E$) and activation energies ($E_a$) for breaking a C−H bond of ethylene, and the positions of the Pd-d band centers with respect to the Fermi levels ($E_{Pd-d}$) on different ensemble of the PdAu(001) bimetallic surface as well as on the clean Pd(001) surfaces. All energies are given in the unit of eV.

| | $Pd_1$ | $Pd_{2,SN}$ | $Pd_{4,SN}$ | $Pd_{2,FN}$ | $Pd_3$ | $Pd_5$ | $Pd_8$ | Pd(100) |
|---|---|---|---|---|---|---|---|---|
| $E_{ad,Ethylene}^{[100]}$ | 0.54 | 0.63 | 0.65 | 0.68 | 0.76 | 0.91 | 0.87 | 0.82 |
| $E_{ad,Ethylene}^{[110]}$ | 0.56 | 0.66 | 0.69 | 0.74 | 0.83 | 0.90 | 0.98 | 0.94 |
| $E_{ad,H}$ | 1.90 | 2.20 | 2.29 | 2.25 | 2.53 | 2.59 | 2.62 | 2.62 |
| $E_{ad,vinyl}$ | 2.01 | 2.04 | 2.25 | 2.26 | 2.39 | 2.59 | 2.65 | 2.61 |
| $\Delta E$ | 1.34 | 1.19 | 1.08 | 1.01 | 0.88 | 0.66 | 0.50 | 0.54 |
| $E_a$ | 1.83 | 1.82 | 1.68 | 1.26 | 1.13 | 0.95 | 0.83 | 1.04 |
| $E_{Pd-d}$ | -1.47 | -1.41 | -1.38 | -1.37 | -1.32 | -1.35 | -1.28 | -1.58 |

After the dehydrogenation process, we have vinyl, a univalent chemical radical $CH_2CH$, and the eliminated H atom on the PdAu(100) bimetallic surface. Since both of them need to be accommodated in a small region, the ensemble effect is expected to play a more significant role in the final state. In the optimized geometries, H and vinyl take the Pd-Pd or Pd-Au bridge sites, sketched in Fig.2. The calculated reaction energies, i.e., the differences between the total energies of the final and initial states $(\Delta E = E_{vinyl+H/PdAu} - E_{ethylene/PdAu})$, are also list in Table 1. Clearly, the reaction energies on large Pd ensembles are comparable to those on the Pd(001) surface where ethylene can be easily dehydrogenated. In contrast, the values of $\Delta E$ on Pd monomers and second neighborhoods are twice larger. To give a more direct assessment of the reaction rates, we also provide activation energies for C-H bond breaking, $E_a$, obtained through the CI-NEB simulations. The CI-NEB approach has been successfully applied in theoretical studies of various chemical reactions, including dehydrogenation of ethylene on Pd(111) and Pd/Au(111) [26,27]. Our calculations indicate that the activation energy is 1.04 eV for ethylene dehydrogenation on the clean Pd(001) surface. It is interesting that $E_a$ decreases on the PdAu bimetallic surfaces with large ensembles such as $Pd_8$ and $Pd_5$. Even on $Pd_{2,FN}$ and $Pd_3$, $E_a$ is comparable to that on Pd(001). As a result, Pd first neighborhoods can effectively decompose ethylene. In contrast, both $E_a$ and $\Delta E$ are much higher on Pd monomers and second neighborhoods, indicating the inability of these ensembles in decomposing ethylene under ambient reaction temperature. To elucidate the origin of the difference between Pd first and second neighborhoods, we present in Fig. 2 the energies along the paths of eliminating one H atom from the ethylene molecule on the $Pd_{2,SN}$ and $Pd_8$ ensembles. For both cases, the eliminated H and vinyl take the adjacent Pd-Pd or Pd-Au bridge site in the transition and final states. From the data in Table 1, the adsorption energies for H and vinyl on Pd-Pd bridge site is about 0.4 eV larger than those on the Pd-Au bridge site. Since only Pd-Au bridge site(s) are available around Pd monomers and second neighborhoods, $E_a$ strongly surges on these ensembles and the dehydrogenation is expected to be inhibited.

As was revealed both experimentally and theoretically, the population of the Pd first neighborhoods is very small on annealed PdAu(001). One can thus conclude that ethylene molecules are unlikely to be decomposed on PdAu bimetallic surfaces. This agrees well with the experimental observations of Goodman et al in their (TPD) measurements.<sup>12</sup><span style="color:red">错误！未定义书签。</span> Apparently, VA synthesis on PdAu catalysts occurs through the direct formation of ethyl acetate-like intermediates followed by β-H elimination, rather than via ethylene dehydrogenation precursor state. It is important to point out here that the reaction rate and reaction mechanism are primarily determined by the ensemble effect, whereas the ligand effect plays only a minor role.

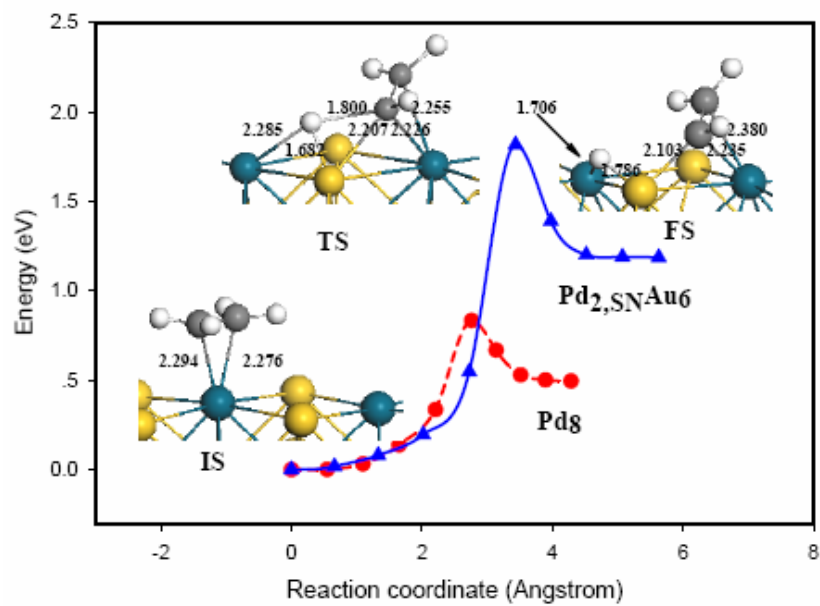

Fig. 2. Reaction energy diagrams and atomic configurations for ethylene dehydrogenation on ensembles of $Pd_{2,SN}$ and $Pd_8$. Bond lengths are given for initial, transition and final states on $Pd_{2,SN}$.

Since calculations for *ΔE* and $E_a$ are computationally demanding, it is instructive to inspect their correlations with $E_{ad,ethylene}$ on different ensembles. Interestingly, both *ΔE* and $E_a$ scales almost linearly with $E_{ad,ethylene}$ in Fig. 3(a). ΔE can be also estimated through adsorption energies of $C_2H_4$, $CH_2CH$ and H as

$$\Delta E = E_{ad,ethylene} - E_{ad,vinyl} - E_{ad,H} + 4.93 \ eV \qquad (2)$$

where *4.93 eV* represents the energy required for breaking a CH bond of ethylene in vacuum. Since H and vinyl take the bridge sites, they effectively explore the ensemble effects. On the other hand

$E_{ad,ethylene}$ is also sensitive to the ensemble underneath with adjustment in adsorption site and orientation. The strong ensemble effect on *ΔE* is somewhat reflected in $E_{ad,ethylene}$. The same argument can be used for understanding the proportionality between $E_a$ and $E_{ad,ethylene}$ since the main difference between the transition and final states is the separation between H and vinyl segments. Although the quality of linear fitting for $E_a \sim E_{ad,ethylene}$ is low, it appears that one can reasonably estimate the catalytic behavior of different ensembles estimated from adsorption energies of reactants alone. This is certainly useful in expediting the process for identifying key

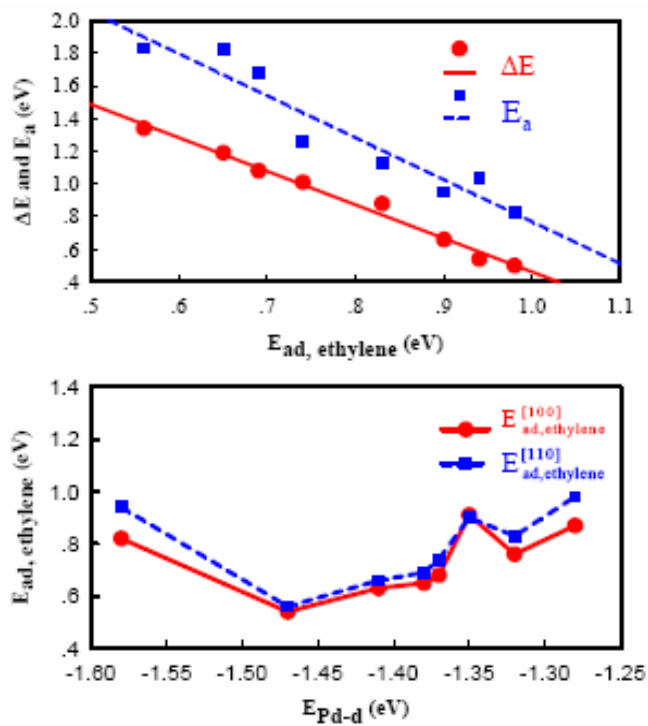

Fig. 3. The correlations between (a) ΔE, $E_a$ and $E_{ad,ethylene}$ and (b) $E_{ad,ethylene}$ and $E_{Pd-d}$.

ensembles for design of efficient catalysts. Z.P. Liu et al also found correlations between adsorption energy of reactants and reaction barriers for several oxidation and hydrogenation <span style="color:red">reactions,</span> but the

trend is opposite to what we observed here.[28]

Since $E_{ad,ethylene}$ is of central importance for the understanding of ethylene dehydrogenation, it is useful to correlate it to more fundamental electronic properties of the substrate. For small molecules such as CO, it was found that $E_{ad}$ scales linearly to the position of Pd-d band center with respect to the Fermi level, $E_{Pd-d}$. Unfortunately, the $E_{ad,ethylene}{\sim}E_{Pd-d}$ plot shown in Fig. 3(b) displays no obvious correlation between these two quantities[29]. This is understandable since $E_{ad,ethylene}$ now strongly depends on the ensemble effect, whereas $E_{Pd-d}$ only reflects the ligand effect. From Table 1, it can be found that $E_{Pd-d}$ changes only slightly for all the ensembles, whereas $E_{ad,ethylene}$ swings in a large range. Therefore, $E_{Pd-d}$ is no longer a good measure for the *local* activity of different ensembles on bimetallic surfaces. For uniform surfaces such as Pd$_8$ and Pd(001), however, the correlation between $E_{ad}$ and $E_{Pd-d}$ appears to hold. For example, $E_{Pd-d}$ of Pd$_8$ is higher by 0.30 eV compared to that of pure Pd(001) surface. Accordingly, $E_{ad,ethylene}$ on the former is larger by 0.05 eV.

The interaction between ethylene and metal surfaces is usually described by the Dewar-Chatt-Duncanson model:[30, 31] the filled $\pi_{CC}$ orbital of ethylene (HOMO) donates charge to the empty d-orbitals of metal substrates, while the occupied d-orbitals of metal back-donate charge to the antibonding $\pi_{CC}^{*}$ orbital of ethylene (LUMO). From the projected density of states (PDOS) of ethylene and Pd-d states of clean and adsorbed PdAu surfaces in Fig. 4, it is obvious that both $\pi_{CC}$ and $\pi_{CC}^{*}$ states are broadened into a wide energy range. Accordingly, the Pd-d states are shifted to the low energy region, compared to the PDOS of the clean PdAu(001) surfaces. The Pd-d band is narrow and well below the Fermi level for the Pd monomer and Pd second neighborhoods. As a result, the electron donation and back-donation become difficult. Meanwhile, the hybridization between Pd-d states and ethylene $\pi_{CC}$ orbitals is weakened, as manifested by the reappearance of the $\pi_{CC}$ peak under the Pd-d band. These factors cause decrease of $E_{ad,ethylene}$ on Pd monomer and second neighborhoods, as seen in Table 1. Although the $\sigma_{CH}^{*}$ peak of ethylene is also affected by the substrate, the key factor that produces high activation energies for hydrogenation on Pd second neighborhoods is the need to place one CH$_2$ fragment on Au in the transition state.

To conclude, first-principles calculations indicate that the dehydrogenation of ethylene is unlikely to occur on the PdAu(100) bimetallic surfaces. This mainly stems from the inability of the popular Pd second ensembles to simultaneously provide strong attraction for vinyl and H. It is clear that the formation of VA does not occur through the direct coupling of vinyl and acetate, but rather through the direct combination of ethylene and acetate followed by eliminating one H atom from the intermediate. The ensemble effects are extremely important for the adsorption, decomposition and chemical reaction of large molecules on bimetallic surfaces. Interestingly, we found a correlation between activation energy and adsorption energy, which allow quick estimation of reaction rates without calculations for the transition and final states. Furthermore, since the ensemble effects play a dominant role, the center of Pd-d band cannot measure the surface activity.

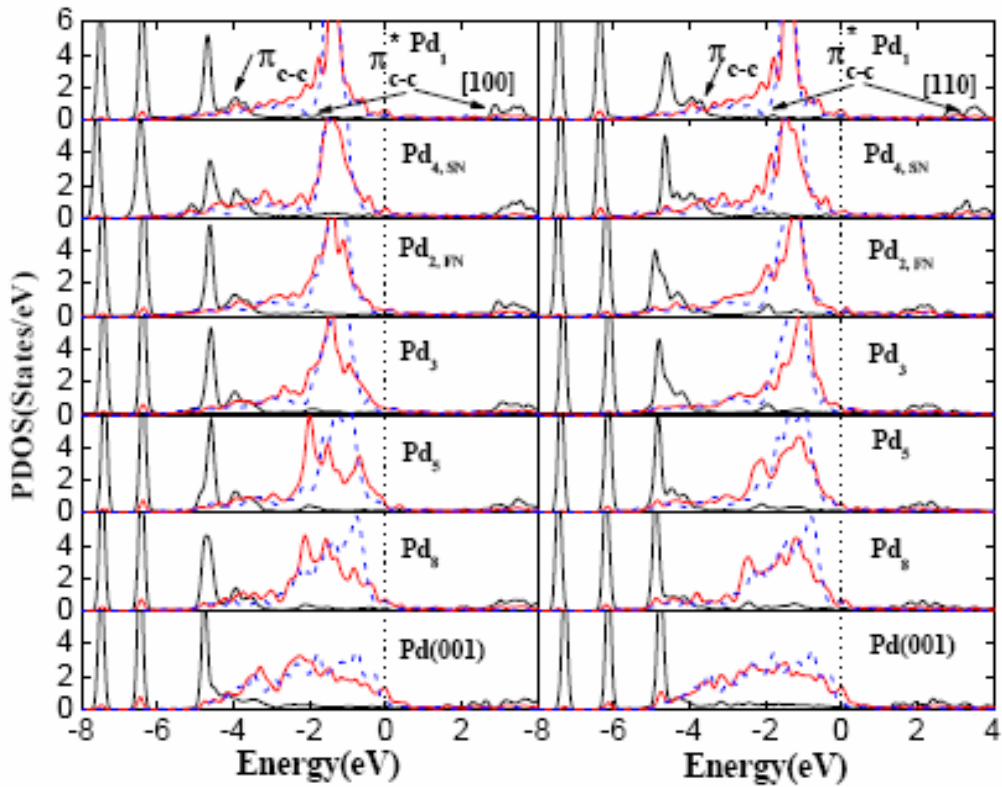

FIG. 4: The PDOS of ethylene (black lines) and Pd-d states (red lines) for $C_2H_4$/PdAu(100). Pd-d states of the clean PdAu(001) surface are plotted with the blue-dashed line.

RW acknowledges simulative discussions with Prof. D.W. Goodman. Work was supported by the DOE-BES (grant No: DE-FG02-04ER15611). XG was supported by the NSF of China, the national program for the basic research and research program of Shanghai. Calculations are performed on supercomputers in the NERSC.

**References:**

1  J. H. Sinfelt, Bimetallic Catalysis: Discoveries, Concepts and Applications, Wiley, New York, 1983.

2  R. W. J. Scott, C. Sivadinarayana, O. M. Wilson, Z. Yan, D. W. Goodman, and R. M. Crooks, J. Am. Chem. Soc. **127**, 1380 (2005).

3  Y. L. Yang, et al. Catal. Commun. **7**, 281 (2006).

4  J. K. Edwards et al., J. Catal. **236**, 69 (2005).

5  D. I. Enache, et al., Science **311**, 362 (2006).

6  M. S. Chem, D. Kumar, C. W. Yi, and D. W. Goodman, Science **310**, 291 (2005).

7  J. A. Rodriguez and D. W. Goodman, Science **257**, 897 (1992).

8  J. R. Kitchin, J. K. Nørskov, M. A. Barteau, and J. G. Chen, Phys. Rev. Lett. **93**, 156801 (2004).

9  B. Hammer and M. Scheffler, Phys. Rev. Lett. **74**, 3487 (1995).

10  B. Hammer, Y. Morikawa, and J. K. Nørskov, Phys. Rev. Lett. **76**, 2141 (1996).

11  F. Maroun, F. Ozanam, O. M. Magnussen, and R. J. Behm, Science **293**, 1811 (2001).

12  K. Luo, T. Wei, C. W. Yi, S. Axnanda , and D. W. Goodman, J. Phys. Chem. **B**, **109**, 23517 (2005).

13  D.W.Yuan , X.G. Gong , and Ruqian Wu, Phys Rev. Lett. submitted.

14  Y. F. Han, D. Kumar, C. Sivadinarayana, A. Clearfield, and D. W. Goodman, Catal, Lett. **94**, 131(2004).

15  M. Neurock and D. H. Mei, Top. Catal. **20**, 5 (2002).

16  D. Stacchiola, F. Calaza, L. Burkholder, A. W. Schwabacher, M. Neurock, and W. T. Tysoe,  Angew, Chem. Int. Ed. **44**, 4572 (2005).

17  G. Kresse and J. Furthmuller, Phys. Rev. B **54**, 11169 (1996).

18  J. P. Perdew and  Y. Wang, Phys. Rev. B **45**, 13244 (1992).

19  D. Vanderbit, Phys. Rev. B **41**, 7892 (1990).

20  H. J. Monkhorst and J. D. Pack, Phys. Rev. B **13** 5188 (1976)

21  G. Schenter, G. Mills, and H. Jnsson, J. Chem. Phys. **101**, 8964(1994).

22  G. Mills, H. Jónsson, and G. Schenter, Surf. Sci. **324**, 305 (1995).

23  G. Henkelman and H. Jónsson, J. Chem. Phys. **113**, 9978(2000).

24  G. Henkelman, B.P. Uberuaga, and H. Jónsson, J. Chem. Phys. **113**, 9901 (2000).

25  Q. Ge, and M. Neurock, Chem. Phys. Lett. **358**, 377(2002).

26  V. Pallassana, M. Neurock, V. S. Lusvardi, J. J. Lerou, D. D. Kragten, and R. A. van Santen, J. Phys. Chem. B, **106**, 1656 (2002).

27  V. Pallassana and M. Neurock, J. Catal. **191**, 301 (2000).

28  Z. P. Liu and P. Hu, J. Chem. Phys. **115**, 4977 (2001).

29  Technically, it is difficult to obtain accurate results for $E_{Pd-d}$ since the d-shell is open and the partial d-DOS has a long tail in the unoccupied region as shown in Fig. 4. Actually, different authors reported diverse results of $E_{Pd-d}$ for the same system, eg., from -2.01 eV to -1.46 eV for the Pd(111) surface (see, Ref. 27 and R. Hirschl and J. Hafner, Surf. Sci. **498**, 21(2002) and A. Roudgar, and A. Groβ, Surf. Sci. **559**, L180(2004)). In the present calculations, we terminate the upper limit of energy integrals up to 4.0 eV above the Fermi level for the determination of $E_{Pd-d}$.

30  F. Zaera, Chem. Rev. **95**, 2651 (1995).

31  M. J. S. Dewar and G. P. Ford, J. Am. Chem. Soc. **101**, 783 (1979).